\documentclass[prl,twocolumn,superscriptaddress]{revtex4}
\usepackage{graphicx,amsmath,amssymb,bm}

\def\be{\begin{equation}}
\def\ee{\end{equation}}
\def\ba{\begin{eqnarray}}
\def\ea{\end{eqnarray}}
\def\bas{\begin{eqnarray*}}
\def\eas{\end{eqnarray*}}

\begin{document}

\title{{\it Ab-initio} computation of the $^{17}$F proton-halo state
  and resonances in $A=17$ nuclei}

\author{G.~Hagen}

\affiliation{Physics Division, Oak Ridge National Laboratory, Oak
  Ridge, TN 37831, USA}

\author{T.~Papenbrock}

\affiliation{Department of Physics and Astronomy, University of
  Tennessee, Knoxville, TN 37996, USA}

\affiliation{Physics Division, Oak Ridge National Laboratory, Oak
  Ridge, TN 37831, USA}

\author{M.~Hjorth-Jensen}

\affiliation{Department of Physics and Center of Mathematics for
  Applications, University of Oslo, N-0316 Oslo, Norway}

\begin{abstract}
  We perform coupled-cluster calculations of the energies and
  lifetimes of single-particle states around the doubly magic nucleus
  $^{16}$O based on chiral nucleon-nucleon interactions at
  next-to-next-to-next-to-leading order. To incorporate effects from
  the scattering continuum, we solve the coupled-cluster equations with
  a Gamow-Hartree-Fock basis.  Our calculations for the
  $J^\pi={1/2}^+$ proton-halo state in $^{17}$F and the 
  ${1/2}^+$ state in $^{17}$O agree well with experiment, while the
  calculated spin-orbit splitting between $d_{5/2}$ and $d_{3/2}$
  states is too small due to the lack of three-nucleon forces. We find
  that continuum effects yield a significant amount of additional
  binding energy for the ${1/2}^+$ and ${3/2}^+$ states in $^{17}$O
  and $^{17}$F.
\end{abstract}


\maketitle 

\emph{Introduction.}  Halo nuclei~\cite{Tanihata}, i.e. very fragile
nuclear systems with a halo consisting of one or more weakly bound
nucleons, are fascinating objects.  Atomic nuclei with halo ground
states exist at the fringes of nuclear existence close to the drip
lines.  Well-known examples are the two-neutron halo nuclei $^{6}$He,
and $^{11}$Li, the proton-halo nucleus $^8$B, and the two-proton halo
nucleus $^{17}$Ne; see Ref.~\cite{Jonson} for a recent review. Halo
states can also exist as excited states of nuclei with well-bound
ground states.  Halo nuclei are difficult to study experimentally due
to their feeble nature and the often small production cross sections.
They also provide theory with a formidable challenge since the
proximity of the continuum introduces a very large number of degrees
of freedom. In recent years, several theoretical approaches have been
implemented and developed that include continuum effects and enable
theorists to describe weakly bound states, nuclear halos, and unbound
resonances~\cite{Bennaceur00,MichelBetan,Volya05,michel07,Nol07,quaglioni08}.

The $A=17$ neighbors around $^{16}$O are particularly interesting and
significant nuclei.  First, the ${3/2}^+$ and ${1/2}^+$ states in
$^{17}$F are bound by only 600~keV and 105~keV, respectively, making
the latter a proton-halo state. This state and astrophysically
relevant reactions such as the $^{17}$F(p,$\gamma$)$^{18}$Ne
reaction~\cite{Bardayan} have been understood within the shell model
embedded in the continuum~\cite{Bennaceur00}, but an {\it ab-initio}
description is not yet available.  Second, the ground and excited
states in $^{17}$F and $^{17}$O determine the single-particle energies
of proton and neutron states with respect to the doubly magic nucleus
$^{16}$O, respectively. These energies are basic ingredients of the
nuclear shell model, and they are also key for the understanding of
the evolution of shell structure in the fluorine and oxygen isotopes
\cite{Sorlin08}. Recent theoretical efforts aim at {\it ab-initio}
shell-model calculations with a core for $sd$-shell
nuclei~\cite{Lis08}. The {\it ab-initio} computation of
single-particle energies in $^{17}$O and $^{17}$F is one necessary
ingredient for such an approach. Finally, the {\it ab-initio} approach
to the proton-halo state in $^{17}$F and the $3/2^+$ resonances in
$^{17}$O and $^{17}$F provides us with an ambitious testing ground for
the employed method, the high-precision potentials, and the role of
three-nucleon forces.

In this Letter, we present an {\it ab-initio} calculation of low-lying
states of the mirror nuclei $^{17}$O and $^{17}$F. The coupled-cluster
method~\cite{CoeCizKue} is ideally suited for this endeavor, as it is
a most efficient approximation for the computation of ground states of
doubly magic nuclei, and the single-particle states in odd-mass
neighbors can be computed within the equation-of-motion techniques
\cite{Gour06}.  For the inclusion of continuum effects, we employ the
Berggren~\cite{Berggren1968} single-particle basis of the Gamow shell
model \cite{MichelBetan}, i.e. the model space consists of bound-,
resonant-, and continuum scattering states. Our calculations employ
the chiral nucleon-nucleon interaction at next-to-next-to-next-to
leading order (N$^3$LO) by Machleidt and Entem~\cite{N3LO}.  

\emph{Interaction and model space.} 
We employ the intrinsic nuclear Hamiltonian 
\ba
\label{ham}
\nonumber
\hat{H} &=& \hat{T}-\hat{T}_{\rm cm} +\hat{V}\\
&=& \sum_{1\le i<j\le A} {(\vec{p}_i-\vec{p}_j)^2\over 2mA} +\hat{V} \ .
\ea
Here, $T$ and $T_{\rm cm}$ denote the kinetic energy and the kinetic
energy of the center-of-mass coordinate, respectively, and $V$ denotes
the chiral nucleon-nucleon interaction by Entem and
Machleidt~\cite{N3LO} at N$^3$LO.  

As some of the states we seek to compute are resonances or loosely
bound halo states, we need to take into account continuum effects.
For this purpose we use a Berggren representation \cite{Berggren1968}
for the proton and neutron $s_{1/2}$, $d_{3/2}$, and $d_{5/2}$ partial
waves. The Berggren representation is a generalization of the usual
completeness relation to the complex energy plane, so that bound-,
resonant-, and non-resonant continuum states are treated on an equal
footing. The Berggren ensemble has been successfully used within the
Gamow shell model~\cite{MichelBetan} (see Ref.~\cite{michel09} for a
recent review), and in \emph{ab-initio} coupled-cluster calculations
of energies and lifetimes of the helium isotopes~\cite{hagen07}.  In
constructing the single-particle Berggren basis, we follow the
procedure outlined in Ref.~\cite{hagen3}. We diagonalize a one-body
Hamiltonian with a spherical Woods-Saxon potential in a spherical-wave
basis defined on a discretized contour $L^+_2$ in the complex momentum
plane. We employ a total of 30 Gauss-Legendre mesh points along the
contour for each of the $s_{1/2}$, $d_{3/2}$, and $d_{5/2}$ partial
waves. Our converged calculations are independent of the choice of
contour, and we checked that 30 mesh points is sufficient to reach
satisfactory converged results for the calculated energies and
lifetimes of the states we consider in this work. For all other
partial waves, the basis functions are those of the spherical harmonic
oscillator.

\emph{Method.}  The computation of the ground and excited states in
$^{17}$O and $^{17}$F is a three-step procedure within coupled-cluster
theory. First, we employ the intrinsic Hamiltonian~(\ref{ham}) and
compute the ground-state energy $E_0$ of $^{16}$O. This yields a
precise reference value relative to which the single-particle energies
will be determined. In the second step, we compute the ground-state
energy $E_0^*$ and corresponding cluster amplitudes for a
``mass-shifted'' nucleus $^{16}$O, where the mass shift $m\to m'=m
(A+1)/A$ in the intrinsic Hamiltonian~(\ref{ham}) ensures that the
correct kinetic energy of the center-of-mass is utilized in the third
step. In the third step, we act with an effective one-particle
creation operator (consisting of superpositions of one-particle and
two-particle-one-hole operators) onto the mass-shifted 
ground state of $^{16}$O. This yields the energies $E_\mu =
E_0^*+\omega_\mu$ of the states with spin and parity
$\mu=1/2^+, 3/2^+, 5/2^+$ in the $A=17$ nucleus of interest. The
difference between these energies and the ground-state energy of
$^{16}$O are the single-particle energies $E_{\rm
  sp}^{(\mu)}$, i.e. $E_{\rm sp}^{(\mu)} = \omega_\mu+E_0^*-E_0$. We
briefly describe the three steps in more detail.

In coupled-cluster theory~\cite{CoeCizKue,Dean04,Bar07}, one computes
the similarity-transformed Hamiltonian $\overline{H}=e^{-T}He^T$ for
the closed-shell nucleus $^{16}$O. Here, $T$ is a sum of particle-hole
cluster operators $T = \sum_{k=1}^A T_k$. 
The $k$-particle $k$-hole ($k$p-$k$h) cluster operator
\begin{equation}
T_k =
\frac{1}{(k!)^2}  t_{i_1\ldots i_k}^{a_1\ldots a_k}
\hat{a}^\dagger_{a_1}\ldots\hat{a}^\dagger_{a_k}
\hat{a}_{i_k}\ldots\hat{a}_{i_1} \ .
\end{equation}
is defined with respect to the Hartree-Fock reference state
$|\phi_0\rangle$. Here, and in the following, we sum over repeated
indices. The labels $i, j, k,\ldots$ ($a,b,c\ldots$) denote occupied
(unoccupied) single-particle orbitals. The operators $\hat{a}_p$
($\hat{a}^\dagger_p$) annihilate (create) a fermion in orbital $p$. In
practice, we truncate the cluster expansion by setting $T_a=0$ for
$a>3$, and treat the triples cluster $T_3$ in the $\Lambda$CCSD(T)
approximation~\cite{KuchTaube}.  The unknown cluster amplitudes
$t_i^a$ and $t_{ij}^{ab}$ are determined from the condition that the
similarity-transformed Hamiltonian $\overline{H}$ has no 1p-1h
excitations and no 2p-2h excitations, respectively, from its
Hartree-Fock reference state. The ground-state energy is the
expectation value of $\overline{H}$ in the Hartree-Fock reference,
with small corrections due to the approximate inclusion of triples
added. This approach is used for the computation of the ground-state
energies $E_0$ and $E_0^*$ of $^{16}$O and the ``mass-shifted''
$^{16}$O, respectively.  We employ the coupled-cluster method in an
angular-momentum coupled scheme~\cite{Hag08,Hag09}. This allows us to
obtain well-converged results for ``bare'' interactions from chiral
effective field theory (EFT) in large model spaces consisting of 
15-20 oscillator shells.

We wish to study the low-lying states in $^{17}$O and $^{17}$F. These
nuclei differ by an additional neutron or proton from the doubly magic
$^{16}$O.  Excited states (with dominant single-particle character)
can be obtained from the ground state of the ``mass-shifted'' $^{16}$O
by action of the excitation operator
\be 
R_\mu = r^aa_a^{\dagger} + {1\over
  2} r^{ab}_j a^{\dagger}_a  a^{\dagger}_b  a_j \ . 
\label{eq:paeqn1}
\ee
Here, it is understood that the annihilation and creation operators on
the right-hand side of Eq.~(\ref{eq:paeqn1}) are coupled to the spin and
parity $\mu$ of the excited proton and neutron states that 
we seek to compute, respectively. This is the
particle-attached equation-of-motion coupled-cluster method
with singles- and doubles excitations (PA-EOM-CCSD), 
see e.g. Refs. \cite{Hirata00,Bar07}. The unknowns
$r^a$ and $r_i^{ab}$ and the excitation energies $\omega_\mu$ relative
to the ground-state energy of the mass-shifted $^{16}$O are obtained
from solving the eigenvalue problem
\begin{equation}
\left[ \overline{H}, R_\mu \right] \vert \phi_0 \rangle = \omega_\mu
R_\mu \vert \phi_0 \rangle \ .
\label{eq:paeqn2}
\end{equation}

\emph{Results.}  We perform a Hartree-Fock (HF) calculation for
$^{16}$O and obtain the reference state $\vert\phi_0\rangle$. In order
to assess the role of coupling to the scattering continuum, we present
results for the $A=17$ system, starting from a Hartree-Fock basis
derived from a Harmonic Oscillator basis (OHF) and a Woods-Saxon
Berggren basis which is the Gamow-Hartree-Fock basis (GHF) \cite{michel09}.  
For well-bound nuclei such as $^{16}$O,
the coupling to continuum degrees of freedom is negligible. The
ground-state energy of $^{16}$O differs by less than 1~keV in the OHF
and GHF basis within both the CCSD and the $\Lambda$CCSD(T)
approximation.  We found well-converged results for the ground state
of $^{16}$O in 15 major oscillator shells, and the energy varies by
less than 0.5MeV for 26~MeV$\le\hbar\omega\le$~36~MeV.  (See
Refs.~\cite{Hag08,Hag09} for convergence details.)  At the
energy minimum $\hbar\omega=34$~MeV, the ground state energy of
$^{16}$O is $-107.6$MeV in the CCSD, and $-120.9$MeV in the
$\Lambda$CCSD(T) approximation.

\begin{figure}[htbp]
\includegraphics[width=0.45\textwidth,clip=]{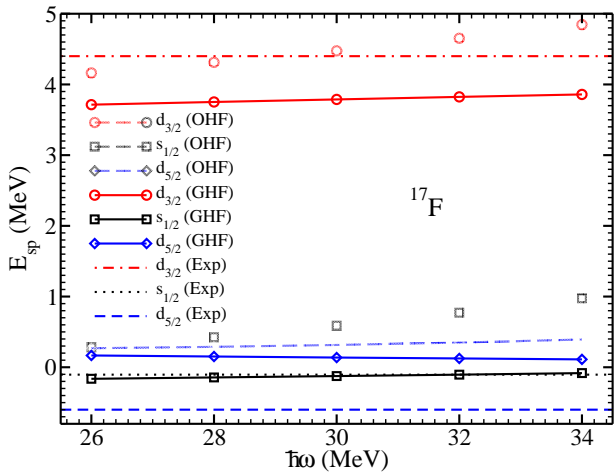}
\caption{(Color online) Low-lying single-particle states in $^{17}$F
  relative to the $^{16}$O ground-state energy as a function of
  oscillator frequency $\hbar\omega$. The data points connected by
  dashed and solid lines employ an oscillator basis (OHF) and a Berggren
  basis (GHF), respectively. The horizontal lines are experimental data.}
\label{fig:fig1}
\end{figure}

Figure~\ref{fig:fig1} shows our PA-EOMCCSD results for the ${1/2}^+$,
${3/2}^+$, and ${5/2}^+$ single-particle energies $E_{\rm sp}$ in
$^{17}$F as a function of $\hbar\omega$.  The data points connected by
dashed and solid lines show the coupled-cluster results obtained in
the OHF basis and the GHF basis, respectively. The horizontal lines
show the experimental single-particle energies. The underlying model
space includes 17 major oscillator shells, in addition to 30
Woods-Saxon Berggren states for each of the $s_{1/2}, d_{3/2}$, and
$d_{5/2}$ partial waves. The results obtained in the GHF basis exhibit
a very weak dependence on the oscillator frequency while this
dependence is stronger for the OHF basis. In particular, the energies
of the $d_{3/2}$ and $s_{1/2}$ in the OHF basis increase with
increasing frequency of the model space.  The ${3/2}^+$ states in
$^{17}$O and $^{17}$F are well-known resonances, and an oscillator
basis is clearly not appropriate to describe these states.  For the
${5/2}^+$ states, we find a much weaker effect from the continuum.
This is expected since the $l=2$ centrifugal barrier localizes this
state inside the barrier and reduces the coupling with the
external scattering continuum.

The coupling with the scattering continuum has a significant effect on
the ${1/2}^+$ and ${3/2}^+$ states of $^{17}$F and $^{17}$O. Our
calculations using a GHF basis yields $\sim 1.0$~MeV in additional
binding energy for these states compared to our calculations with the
OHF basis. The effect is particularly strong in the case of the
${1/2}^+$ proton halo state in $^{17}$F, which is not even bound in
the OHF basis. Similar continuum coupling effects were found for the
${1/2}^+$ halo state in $^{11}$Be~\cite{quaglioni08} and in the
low-lying states of the fluorine and oxygen
isotopes~\cite{Bennaceur00,Michel10}.  The lack of a centrifugal
barrier and the very weak binding yield a proton-halo (with a
root-mean-square radius of $r_{\rm rms} = 5.333$fm \cite{Morlock97})
that is difficult to capture in the oscillator basis.  Our calculated
binding energy for this state agrees remarkably well with the
experimental value of $105$~keV. This finding deserves further
analysis, and we need to estimate the effects of the omitted
three-nucleon forces.

\begin{figure}[htbp]
\includegraphics[width=0.45\textwidth,clip=]{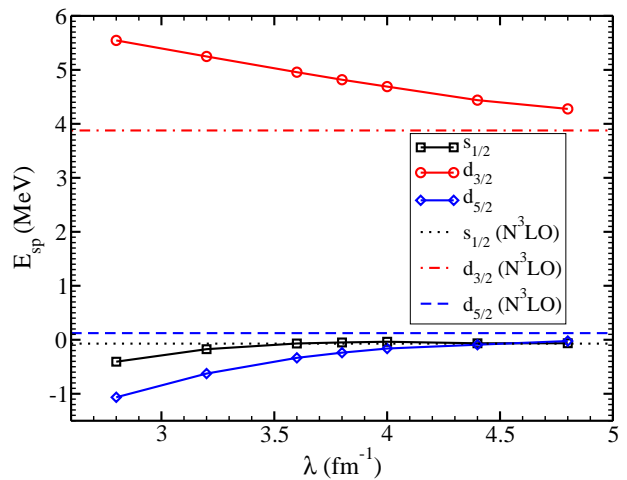}
\caption{(Color online) Single-particle energies of the $s_{1/2}$,
  $d_{3/2}$, and $d_{5/2}$ states in $^{17}$F (squares, circles, and
  diamonds, respectively) and the results for a ``bare'' N$^3$LO
  interaction (dotted, dashed, and dashed-dotted lines, respectively)
  as a function of the high-momentum cutoff $\lambda$.}
\label{fig:fig2}
\end{figure}

Within chiral EFT, the leading three-nucleon forces consist of a
long-range two-pion exchange, a midrange one-pion exchange, and the
short-range three-nucleon contact interaction~\cite{Epel}.
Three-nucleon forces are expected to yield additional binding of the
order of 0.5~MeV per nucleon~\cite{Hag09}. The effect of three-nucleon
forces on energy differences is more subtle. Within a calculation
based on two-nucleon forces we can, however, probe the effect of the
three-body contact by a variation of the ultraviolet cutoff $\lambda$.
Decreasing the cutoff employed in the construction of the chiral
interactions renormalizes the two-nucleon interaction and generates
short-ranged three nucleon forces~\cite{NBS}. We employ the similarity
renormalization group (SRG) \cite{bogner2007} for the generation of
interactions with a cutoff $\lambda$, and study the evolution of the
excited states in $^{17}$F as the cutoff is varied.
Figure~\ref{fig:fig2} shows that the spin-orbit splitting between the
$d_{3/2}$ and $d_{5/2}$ orbitals increases with decreasing cutoff.
However, the $s_{1/2}$ state remains virtually unchanged as the cutoff
is lowered to $\lambda\approx 3.2$~fm$^{-1}$.  This is not unexpected
since the structure of the dilute ${1/2}^+$ halo state is dominated by
long-ranged forces, and the SRG interactions only change the
short-range contributions. Thus, our result for the proton-halo state
in $^{17}$F is insensitive to short-range three-nucleon forces.

Let us also comment on the center-of-mass motion.
Ref.~\cite{Hagen_CoM} demonstrates that the intrinsic
Hamiltonian~(\ref{ham}) yields a coupled-cluster wave function that
factorizes to a very good approximation into an intrinsic part and a
Gaussian for the center-of-mass motion. At the cutoff
$\lambda=2.8$~fm$^{-1}$ we checked that this is true for the low-lying
states in the $A=17$ nuclei in a wide range of oscillator frequencies.

Table~\ref{tab1} summarizes our PA-EOMCCSD results for the      
${1/2}^+$, ${3/2}^+$,  and ${5/2}^+$ states in $^{17}$O and $^{17}$F, and 
compares with experiment. The oscillator frequency is $\hbar\omega=34$~MeV, 
which corresponds to the energy minimum of the CCSD
and $\Lambda$CCSD(T) ground-state energies of $^{16}$O. We also 
show the spin-orbit splitting between the $d_{5/2}$ and $d_{3/2}$ 
single-particle states.

\begin{table}[h]          
\begin{tabular}{|l||l|l|l||l|l|l|}\hline            
\multicolumn{1}{|c||}{}   & \multicolumn{3}{c||}{ $^{17}$O}  &            
\multicolumn{3}{|c|}{ $^{17}$F}  \\ \hline            
& ${1/2}^+$ & ${5/2}^+$ &            
$\mathrm{E}_{\mathrm{so}}$ &            
${1/2}^+$ & ${5/2}^+$ &            
$\mathrm{E}_{\mathrm{so}}$  \\\hline            
GHF         & -2.8   & -3.2   & 4.3   & -0.082 &  0.11  &  3.7  \\\hline     
Exp.        & -3.272 & -4.143 & 5.084 & -0.105 & -0.600 &  5.000   \\ \hline    
\end{tabular}        
\caption{Single-particle energies of the ${1/2}^+$ 
  and ${5/2}^+$ states, and the spin-orbit splitting 
  $\mathrm{E}_{\mathrm{so}} (d_{3/2}$-$d_{5/2})$ (in units of MeV) 
  in $^{17}$O and  $^{17}$F calculated in a Berggren (Gamow) 
basis (GHF), and the comparison to experiment \cite{Tilley93}.} 
\label{tab1}  
\end{table}

Let us also check the consistency of the PA-EOM-CCSD approximation. By
normalizing the excitation amplitudes in Eq.~(\ref{eq:paeqn1}) to one,
we can compare the norms of the $r^{a}$ and $r^{ab}_{j}$ amplitudes
and get a measure of the one-particle structure of the $A=17$ states.
For the low-lying ${3/2}^+$, ${1/2}^+$, and ${5/2}^+$ states in
$^{17}$F, we find $\vert r^{a}r^{a}\vert = 0.87$, $\vert r^{a}
r^{a}\vert = 0.92$, and $\vert r^{a} r^{a}\vert = 0.87$, respectively.
We find similar norms for the states in $^{17}$O. This clearly shows
that these states are dominated by one-particle excitations from the
$^{16}$O ground state, and the PA-EOM-CCSD approximation is known to
perform very well in this case \cite{Hirata00}.

Within the GHF basis, we obtain a width for the resonance states.
Table~\ref{tab2} shows the calculated energy and width of the
$d_{3/2}$ single-particle resonance in $^{17}$O and $^{17}$F in a
model space with $\hbar\omega=34$~MeV. The energy of the $d_{3/2}$
single-particle state in $^{17}$O compares very well with experiment
while in $^{17}$F it is within 0.5~MeV. The calculated widths are very
reasonable compared to the experimental values, and represent the
first \emph{ab-initio} calculation of resonance in an $A=17$ nucleus.

\begin{table}[h]          
\begin{tabular}{|l|l|l|l|l|}\hline            
\multicolumn{1}{|c|}{}   & \multicolumn{2}{c|}{$^{17}$O ${3/2}^+$}  & \multicolumn{2}{c|}{$^{17}$F ${3/2}^+$} \\ \hline 
                &$E_{\rm sp}$&$\Gamma$&$E_{\rm sp}$  & $\Gamma$ \\ \hline  
This work       &  1.1   & 0.014  &  3.9     &  1.0     \\            
Experiment      &  0.942 & 0.096  &  4.399   &  1.530   \\ \hline          
\end{tabular}        
\caption{Computed ${3/2}^+$ single-particle resonance energies 
in $^{17}$O and
  $^{17}$F compared to data \cite{Tilley93}. 
  The real part $E_{\rm sp}={\rm Re}[E]$, and
  the width $\Gamma = 2{\rm Im}[E]$ are given in units of
  MeV.}
\label{tab2}  
\end{table}

\emph{Conclusions.}  We performed \emph{ab-initio} coupled-cluster
calculations of the energy and lifetimes of the low-lying ${1/2}^+$,
${3/2}^+$, and ${5/2}^+$ states in $^{17}$O and $^{17}$F employing
chiral nucleon-nucleon interactions and a Berggren single-particle
basis. The single-particle energy of the ${1/2}^+$ proton halo state
in $^{17}$F agrees well with the experiment, and we checked by cutoff
variation that this result is not affected by short-ranged
three-nucleon forces. We find a reduced $d_{3/2}$-$d_{5/2}$ spin-orbit
splitting compared to experiment, and confirmed via cutoff variation
that this is sensitive to short-ranged three-nucleon forces. The
lifetimes of the ${3/2}^+$ resonances in $^{17}$F and $^{17}$O agree
reasonably well with experimental data. Our calculations also show
that the inclusion of continuum effects is necessary for a proper
description of the studied single-particle states.

We thank W. Nazarewicz for useful discussions. 
This work was supported by the U.S. Department of Energy,
under Grant No.\ DE-FG02-96ER40963 (University of Tennessee), 
and under DE-FC02-07ER41457 (UNEDF SciDAC Collaboration). 
This research used computational resources of the
National Center for Computational Sciences at Oak Ridge National Laboratory.

\end{document}